\documentclass[journal]{IEEEtran}

\usepackage{bibentry}  
\usepackage{xcolor,soul,framed} 

\colorlet{shadecolor}{yellow}
\usepackage[pdftex]{graphicx}
\graphicspath{{../pdf/}{../jpeg/}}
\DeclareGraphicsExtensions{.pdf,.jpeg,.png}

\usepackage[cmex10]{amsmath}
\usepackage{array}
\usepackage{mdwmath}
\usepackage{amssymb}
\usepackage{mdwtab}
\usepackage{eqparbox}
\usepackage{url}
\usepackage{subcaption}
\usepackage{amsmath}
\usepackage{graphicx} %
\usepackage{relsize}
\UseRawInputEncoding

\usepackage[pdfstartview=XYZ,
bookmarks=true,
colorlinks=true,
linkcolor=blue,
urlcolor=blue,
citecolor=blue,
pdftex,
bookmarks=true,
linktocpage=true, 
hyperindex=true
]{hyperref}
\usepackage{orcidlink}

\begin{document}
\bstctlcite{IEEEexample:BSTcontrol}
    \title{Experimental Investigation of Drain Noise in High Electron Mobility Transistors: Thermal and Hot Electron Noise}
  \author{Bekari Gabritchidze\orcidlink{0000-0001-6392-0523}, 
      Kieran A. Cleary\orcidlink{0000-0002-8214-8265}, 
       Anthony C. Readhead\orcidlink{0000-0001-9152-961X}, 
       Austin J. Minnich \orcidlink{0000-0002-9671-9540}
  \thanks{This work was supported by the National Science Foundation under Grant No.~$1911220$.(\textit{Corresponding author: A. J. Minnich})} 
  \thanks{B. Gabritchidze is with the Division of Physics, Mathematics and Astronomy, California Institute of Technology, Pasadena, CA, 91125, and also with the Department of Physics, University of Crete, GR-70013 Heraklion, Greece (bekari@caltech.edu)}
  \thanks{K. A. Cleary and A. C. Readhead are with the Division of Physics, Mathematics and Astronomy, California Institute of Technology, Pasadena, CA, 91125.(kcleary@astro.caltech.edu, acr@astro.caltech.edu)}
  \thanks{A. J. Minnich is with Division of Engineering and Applied Science, California Institute of Technology, Pasadena, CA, 91125(aminnich@caltech.edu) }
     }  %




\maketitle

\begin{abstract}

We report the on-wafer characterization of $S$-parameters and microwave noise temperature ($T_{50}$) of discrete  metamorphic InGaAs high electron mobility transistors (mHEMTs) at 40 K and 300 K and over a range of drain-source voltages ($V_{DS}$).  From these data, we extract a small-signal model and the drain (output) noise current power spectral density ($S_{id}$) at each bias and temperature. This procedure enables $S_{id}$ to be obtained while accounting for the variation of small-signal model, noise impedance match, and other parameters under the various conditions. We find that the thermal noise associated with the channel conductance can only account for a portion of the measured output noise. Considering the variation of output noise with physical temperature and bias and prior studies of microwave noise in quantum wells, we hypothesize that a hot electron noise source based on real-space transfer of electrons from the channel to the barrier could account for the remaining portion of $S_{id}$. We  suggest further studies to gain insights into the physical mechanisms. Finally, we calculate that the minimum HEMT noise temperature could be reduced by up to $\sim 50$\% and $\sim 30$\% at cryogenic temperature and room temperature, respectively, if the hot electron noise could be suppressed.

\end{abstract}

\begin{IEEEkeywords}
High electron mobility transistors, cryogenic electronics, microwave noise, low-noise amplifiers, drain temperature, real-space transfer
\end{IEEEkeywords}

%
\IEEEpeerreviewmaketitle

\section{Introduction}




\IEEEPARstart{H}{igh} electron mobility transistors are widely employed in microwave amplifiers due to their low noise characteristics\cite{mimura_early_2002, longhi_technologies_2019, reinhard_ultra-low_2013,  IEEEIMS:Ultra_low_Noise, schwierz_modern_2002}.While significant improvements have been made in their noise and frequency performance in past decades \cite{IEEEIMS:cha_300, Cha_MMIC_BroadB, Deal_Will, thome_67-116-ghz_2021, cuadrado-calle_broadband_2017}, achieving further improvements requires a physics-based understanding of the origin of microwave noise in low noise HEMTs which is lacking. Presently, noise in HEMTs is interpreted with the  Pospieszalski model \cite{IEEEIMS:modelling_Pospieszalski} in which noise is described using  two uncorrelated  noise generators  at the input ($S_{\upsilon g}$) and output ($S_{id}$). The input noise generator, $S_{\upsilon g}$, is described using a noise temperature $T_{g}$ which is assigned to the intrinsic gate resistance, and this noise temperature is generally accepted to be the physical temperature of the gate resistance \cite{pospieszalski_dependence_2017, pospieszalski_interpreting_2010}. This gate resistance thermal noise is significantly larger than the induced gate noise for frequencies far below the cut-off frequency \cite{heymann_guide_2021}. The output noise generator, $S_{id}$, is described by assigning a noise temperature, $T_{d}$, to the intrinsic output conductance $g_{ds}$. $T_{d}$ is generally taken as a fitting parameter.
Despite the utility of the model in interpreting noise measurements on HEMTs, it is unable to provide insight into the physical origin of the output noise. 

Other works have proposed various mechanisms; for instance, thermal noise has been suggested as the origin of channel noise in field effect transistors (FETs) \cite{der_ziel_thermal_1962}
such as HEMTs \cite{van_der_ziel_thermal_1983, van_der_ziel_hot_electrons} and MOSFETs \cite{Klassen_Prins_A, Klassen_Prins_B}. A recent work attributed drain noise to suppressed shot noise \cite{IEEEIMS:hemt_limits_pospieszalski} while others to thermal and suppressed shot noise \cite{ShotNoise_ThermalNoise}.

Another theory attributes drain noise to microwave partition noise arising from real-space transfer (RST) \cite{esho_theory_2022}, a process that has been investigated in early studies of transport and noise in quantum wells \cite{matulionis_qw-shape-dependent_1997, matulionis_hot-electron_2000, hartnagel_microwave_2001, aninkevicius_1994, aninkeviius_1993}. In this mechanism, electrons are heated by the electric field under the gate to physical temperatures exceeding 1000 K, a temperature sufficiently high that some electrons may thermionically emit out of the channel into the barrier. Because the barrier mobility is substantially less than that of the channel, two dissimilar conduction pathways exist from source to drain, creating partition noise \cite{ambrozy_electronic_1983} at the HEMT output. However, which of these mechanisms is the actual origin of drain noise in low-noise HEMTs remains under investigation. 

In this paper, we performed on-wafer $S$-parameter and microwave noise measurements ($T_{50}$) of discrete metamorphic InGaAs mHEMTs at 40 K and 300 K and various $V_{DS}$ using a cryogenic probe station.  At each temperature and bias we extracted a small-signal model (SSM) and noise parameters from the S-parameter and $T_{50}$ measurements, allowing the output current noise PSD  ($S_{id}$) to be extracted  while accounting for changes in small-signal parameters and optimal noise impedance at each condition. We find that the variation of $S_{id}$ with physical temperature is inconsistent with suppressed shot noise as the sole noise source, and further, that $S_{id}$ cannot be explained only by channel thermal noise. Considering the conclusions of prior studies of microwave noise in quantum wells, we propose that the additional hot-electron noise is due to real-space transfer of electrons from the channel to the barrier, and we suggest how this hypothesis could be tested in future studies. Finally, we compute that the minimum noise temperature could be decreased by 50\% and 30\% at cryogenic and room temperature, respectively, if the hot-electron noise were suppressed.


This paper is organized as follows. A brief description of the experimental set up and the modeling is presented in Section~\ref{on_wafer_characterization}.The $S$-parameter and microwave noise characterization and the extracted drain noise current PSD are presented in Section~\ref{experimental_and_modeling_results}. The contributions of thermal noise to drain noise is presented in Section~\ref{physical_model_for_drain_noise}, followed by a discussion of the hot electron part of drain noise and a possible explanation based on RST in Section~\ref{discussions}. Finally, we provide a summary of the paper in Section~\ref{conclusion}. 

\section{On-Wafer Cryogenic Characterization and modeling} \label{on_wafer_characterization}

The $S$-parameters and the microwave noise temperature  of metamorphic InGaAs mHEMTs (OMMIC, D007IH, 4F50, gate length $L_{g} = 70$ nm) were measured using a cryogenic probe station (CPS) \cite{IEEEIMS:russell_cryogenic_cps}. Details of the measurement procedure and the experimental set-up were described earlier  \cite{bekari}. In brief, the $S$-parameter measurements were carried out in the frequency range 1 -- 18 GHz using a vector network analyzer (VNA, Rhode\&Schwartz ZVA50). The system was calibrated by transferring the measurement plane  from the VNA to the tips of the wafer probes (GGB industries, 40A-GSG-100-DP) by through-reflect-match (TRM) calibration on a  CS-5 calibration substrate (GGB Industries) at each physical temperature. 

The microwave noise temperature was measured  using the Y-factor method with a cold attenuator \cite{bekari, IEEEIMS:fernandez_noise_temperature} at a generator impedance of 50 $\Omega$. The CPS was configured for noise measurements from 2$-$18 GHz, however, the measurement bandwidth was chosen from 5$-$15 GHz
to minimize RF probe$-$pad contact time and avoid RF pad damage due to chuck vibrations. A 10 dB cryogenic and a room temperature attenuator (Quantum Microwave) were inserted between the noise source (Keysight, N4002A) and the device under test (DUT). The room-temperature attenuator was connected directly to the noise source (NS) to reduce the change in output reflections from the `on' to `off' state of the NS, while the cryogenic attenuator was connected to the RF input probe contacting the input of the DUT and served as a cold load. After the DUT, a cryogenic amplifier (Cosmic Microwave, CIT1-18) was used after the output probe within the probe station, and a combination of mixer, oscillator, filters, amplifiers and a power sensor were used outside the probe station to further process and measure the incident noise power. A detailed schematic was presented in \cite{bekari}.

An  analysis indicated that the uncertainty in noise temperature arose primarily from uncertainty in the input loss. These losses consist of stainless steel coaxial cable, cryogenic 10 dB attenuator, cryogenic bias tee, and the input RF probe. These losses were characterized in a separate cryogenic dewar at each temperature for which noise measurements were performed. The insertion loss of the input RF probe was measured at room temperature by measuring the return loss with the probe tips open,  but it could not be measured at cryogenic temperatures. Therefore, an uncertainty of 0.1 dB on the input loss was assumed for the RF probe to account for any changes in its insertion loss as $T_{ph}$ is decreased from 300 K to 40 K. Its temperature was assumed to be that of the cryogenic bias tee connected to the RF probe. Based on an uncertainty of 0.1 dB in the input loss, an absolute uncertainty in $T_{50}$ of $\sim 20$\% at $T_{ph}= 40$ K and $\sim 15$\% at $T_{ph}=300$ K  was determined. The repeatability of the $T_{50}$ measurements was $\lesssim 0.5$ K at $T_{ph}= 40$ K and $\sim 1$ K at $T_{ph}=300$ K.

\begin{figure}
\centering
\includegraphics[width=100mm, height=5 cm]{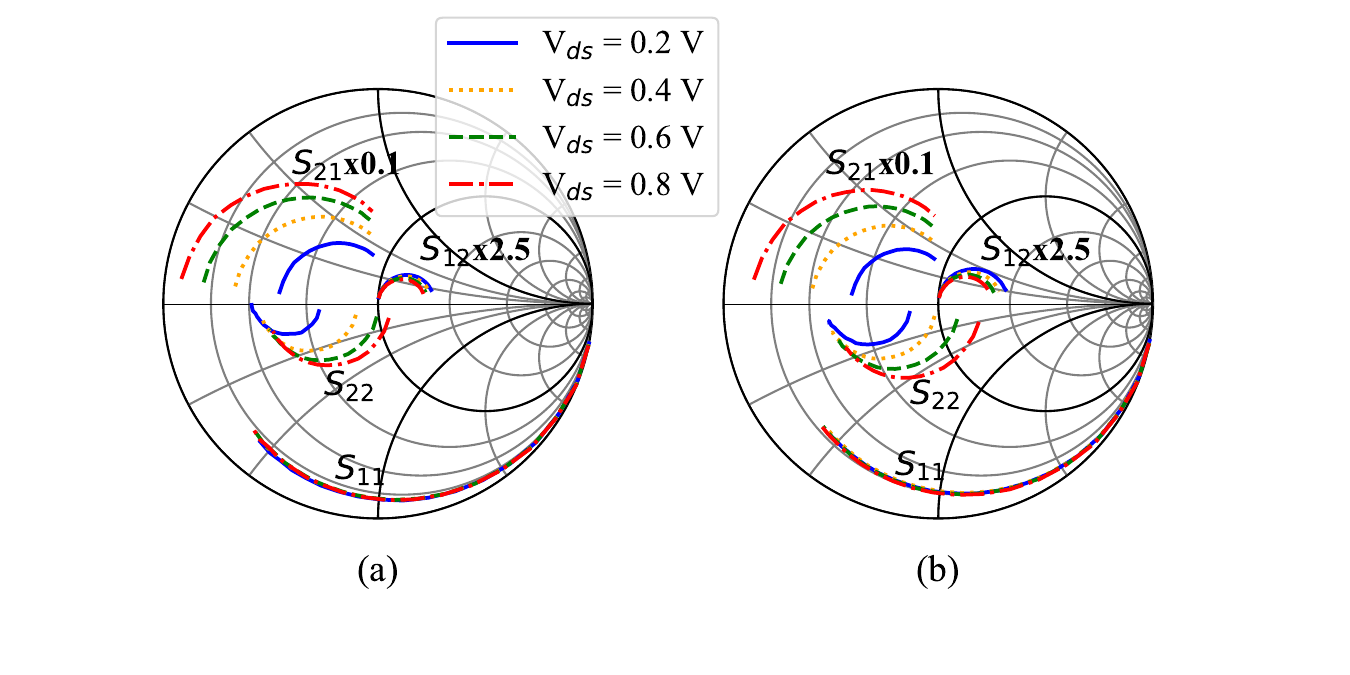}
\caption{ Measured $S$-parameters versus frequency over 1 -- 18 GHz at various $V_{DS}$ and physical temperatures of 40 K (a) and 300 K (b). }
\label{fig:Spars_GaAs}
\setlength{\belowcaptionskip}{-5pt}
\end{figure}

From the $S$-parameter and $T_{50}$ measurements, a 15-element small-signal model (SSM) was developed. First, the parasitics were extracted through cold-FET measurements following \cite{Dambrine_modelling, IEEEIMS:Berroth_FET,hower_current_1973, IEEEIMS:Alt_A, akgiray_new_2013}. Once the parasitics were determined and de-embedded, the intrinsic parameters were evaluated by direct extraction and optimization of the intrinsic parameters. Simulated annealing and quasi-Newton optimization, available in Advanced Design System (ADS, Keysight), were used to minimize a least-square error function following \cite{bekari}. After the optimization, the intrinsic parameters were further fine tuned to improve the agreement between the measured and simulated $S-$parameters. Temperature independence of the extracted parasitic capacitances and inductances was assumed based on \cite{heinz_scalable_2022,IEEEIMS:Alt_A}, and the 300 K values were used at all $T_{ph}$. The parasitic resistances ($R_{g}$, $R_{s}$, $R_{d}$) have been reported to vary $\sim$50\% as $T_{ph}$ is decreased from 300 K to 40 K \cite{IEEEIMS:Alt_A}; however, these resistances are difficult to measure at cryogenic temperatures. We estimated the effect of decreasing parasitic resistances by 50\% in the 40 K SSM on the SSM parameters and $S_{id}$ and found that this change  altered  $g_{m}$ and $R_{ds}$ by $\sim 5$\% and $S_{id}$ by $\sim 6$\%. Therefore, we neglected the temperature dependence of the parasitic resistances. Additionally, the parasitics are bias-independent \cite{IEEEIMS:Alt_A}, so at a given $T_{ph}$ the observed $V_{DS}$ dependent trends are not affected by the assumption of constant parasitic resistance. However, the absolute values of minimum noise temperature ($T_{min}$) with $V_{DS}$ at $T_{ph}=40$ K are overestimated by $\sim 10$\% at 6 GHz when the parasitic resistance at 300 K is used for all temperatures.
SSM parameter extraction was carried out for $V_{DS} \geq 0.1$ V, as in this voltage range the gain exceeded 10 dB and the SSM yielded an average normalized error function \cite{malmkvist_optimization_2008} of $\sim 2$\%. 

Under these assumptions, a noise model based on the Pospieszalski model \cite{IEEEIMS:modelling_Pospieszalski} was obtained by fitting the measured and modeled $T_{50}$ using $S_{id}$ as a fitting parameter. The uncertainties in $S_{id}$ reported as the error bars in the subsequent figures include the uncertainties in access resistances, $g_{ds}$ and $T_{50}$ measurements; however, the total uncertainty in $S_{id}$ is dominated by the uncertainty in our $T_{50}$ measurements.



\begin{figure}
\centering
\includegraphics[width=0.91\columnwidth, height=11.5 cm]{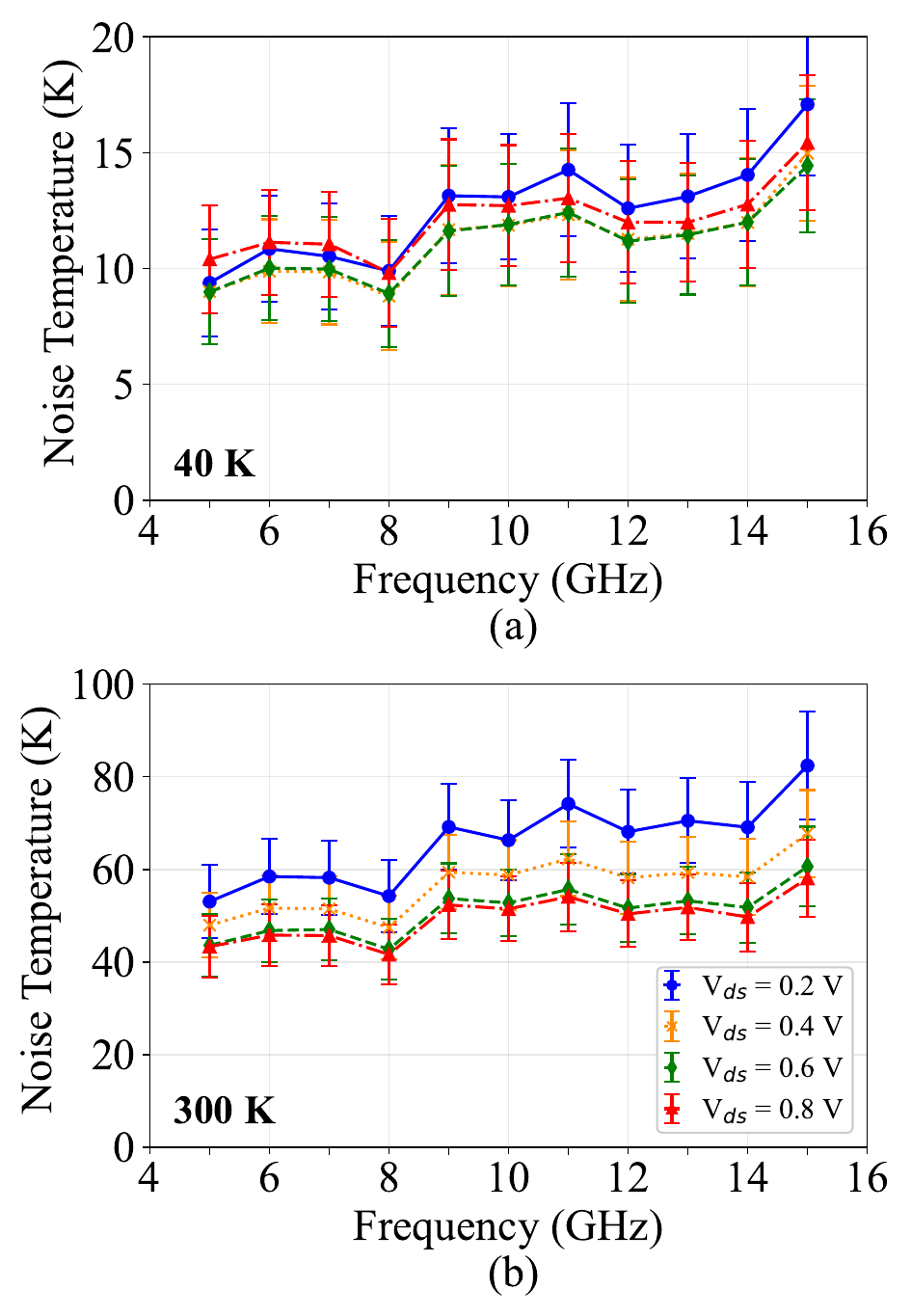}
{
\phantomsubcaption\label{fig:T50_freq_GaAs_40K}
\phantomsubcaption\label{fig:T50_freq_GaAs_300K}
}
\caption{  Measured $T_{50}$ versus frequency at various $V_{DS}$ and physical temperature of 40 K (a) and 300 K (b). The $V_{GS}$ remained constant at all $V_{DS}$ with $V_{GS}=-136$ mV and $V_{GS}=-226$ mV at 40 K and 300 K, respectively.}
\label{fig:T50_GaAs}
\end{figure}

\section{  Results} \label{experimental_and_modeling_results}

Fig.~\ref{fig:Spars_GaAs} shows the measured $S$-parameters at various $V_{DS}$ ranging from 0.2 -- 0.8 V and physical temperatures of 40 K  and 300 K.  At a given $T_{ph}$,  $S_{21}$ and $S_{22}$ exhibit the largest variation with $V_{DS}$, while $S_{12}$ and $S_{11}$ show smaller but observable dependence. In Fig.~\ref{fig:T50_GaAs}, $T_{50}$ versus frequency is shown for various $V_{DS}$ and the two physical temperatures. In both the $S$-parameter and noise measurements, the gate voltage ($V_{GS}$) was kept constant at $V_{GS} = -136$ mV at $T_{ph} = 40$ K, and $V_{GS} = -226$ mV at $T_{ph} = 300$ K,  for all  $V_{DS}$. These $V_{GS}$ were selected so that $I_{DS} = 20$ mA at  $V_{DS} = 0.8$ V and a given $T_{ph}$. At a given $V_{GS}$ and $T_{ph}$, the threshold voltage ($V_{th}$) varied with $V_{DS}$ from $-0.23$ V to $-0.28$ V at 40 K and  $-0.31$ V to $-0.38$ V at 300 K. The $V_{th}$ at each temperature was calculated from the $I_{DS} - V_{GS}$ curves following \cite{schroder_semiconductor_2015}. $V_{DS}$ was varied from 0.1 V to 1.0 V; $V_{DS}$ was restricted to $ \leq 1.0$ V so that gate leakage currents ($I_{G}$) were $I_{G} \lesssim 20$ $\mu$A$/$mm at $T_{ph} = 300$ K, leading to negligible contribution from shot noise \cite{pospieszalski_extremely_2005}. The lowest $T_{ph}$ was limited to 40 K to avoid cryogenic self-heating effects \cite{schleeh_phonon_2015,choi_characterization_2021, ardizzi_self-heating_2022}.


From these measurements, a noise model was developed which permits the extraction of $S_{id}$  at each bias and temperature using the measured $T_{50}$ and S-parameters while accounting for mismatch between $50$ $\Omega$ and optimum noise impedance. Following the CMOS and MOSFET noise literature \cite{excess_noise_CMOS,channel_noise_MOSFETs_A, non_thermal_excess_noise_A,  smit_rf-noise_2014} we plot $S_{id}$ versus $V_{DS}$ at 40 K and 300 K in Figs.~\ref{fig:Td_Vds_GaAs_40K} and ~\ref{fig:Td_Vds_GaAs_300K}, respectively. At 40 K, the $S_{id}$ follows a linear trend at low $V_{DS} \lesssim 0.6$ V and rises rapidly at higher voltages. A similar trend is found at 300 K.


Regarding the rapid increase in $S_{id}$ at high voltages, we note that no indications of non-ideal effects like impact ionization were present in the DC characteristics or $S-$parameters in our devices. The I-V curves (available in \cite[Fig.~3]{bekari}) did not show any kink, while the gate--source current, $I_{GS}$, remained at values  between  -5 $\mu$A$/$mm and -20 $\mu$A$/$mm, far from the bias regime where impact ionization was observed previously in our devices\cite[Fig.~6.7]{akgiray_new_2013}. Additionally, $S_{22}$ remained capacitive for all $V_{DS}$ \cite{IEEEIMS:Ruiz_ImpactIon}.  However, it has been reported based on Monte Carlo simulations \cite{Impact_ionization_and_Sid_A} that microwave noise from impact ionization may occur  even though it may not be observable in I-V and $S-$parameters measurements. Therefore, it is possible that the rapid increase in $S_{id}$ above 0.8 V is due to impact ionization. 
 
The variation of $T_{d}$ with $T_{ph}$ for the present devices has been previously reported in \cite{bekari}. Here, we use those data and the present measurements to obtain $S_{id}$ versus $T_{ph}$ at constant $V_{DS} = 0.6$ V and $I_{DS} = 10$ mA in Fig.~\ref{fig:Td_Te_RST_Vs_Tph_GaAs}. A dependence of $S_{id}$ on $T_{ph}$ is observed. This observation contradicts the predictions of suppressed shot noise theory.  According to this theory, if $I_{DS}$ is kept constant with physical temperature, then the output noise current $S_{id}$ is predicted to be temperature-independent. \cite{IEEEIMS:hemt_limits_pospieszalski, Fano_Factor_Tph_independence, supressed_shotN_excess_noise} This prediction is inconsistent with the measured temperature dependence of the output noise current in Fig.~\ref{fig:T50_Tpred_Tph}. This conclusion agrees with the findings of \cite{InP_DrainNoise_IMS2023, bekari, heinz_scalable_2022, IEEEIMS:murti_Td}. This finding indicates that suppressed shot noise cannot be the sole mechanism contributing to drain noise.


\section{ Contribution of channel thermal noise to output noise } \label{physical_model_for_drain_noise}

\begin{figure}
\centering
\includegraphics[width=0.9\columnwidth] {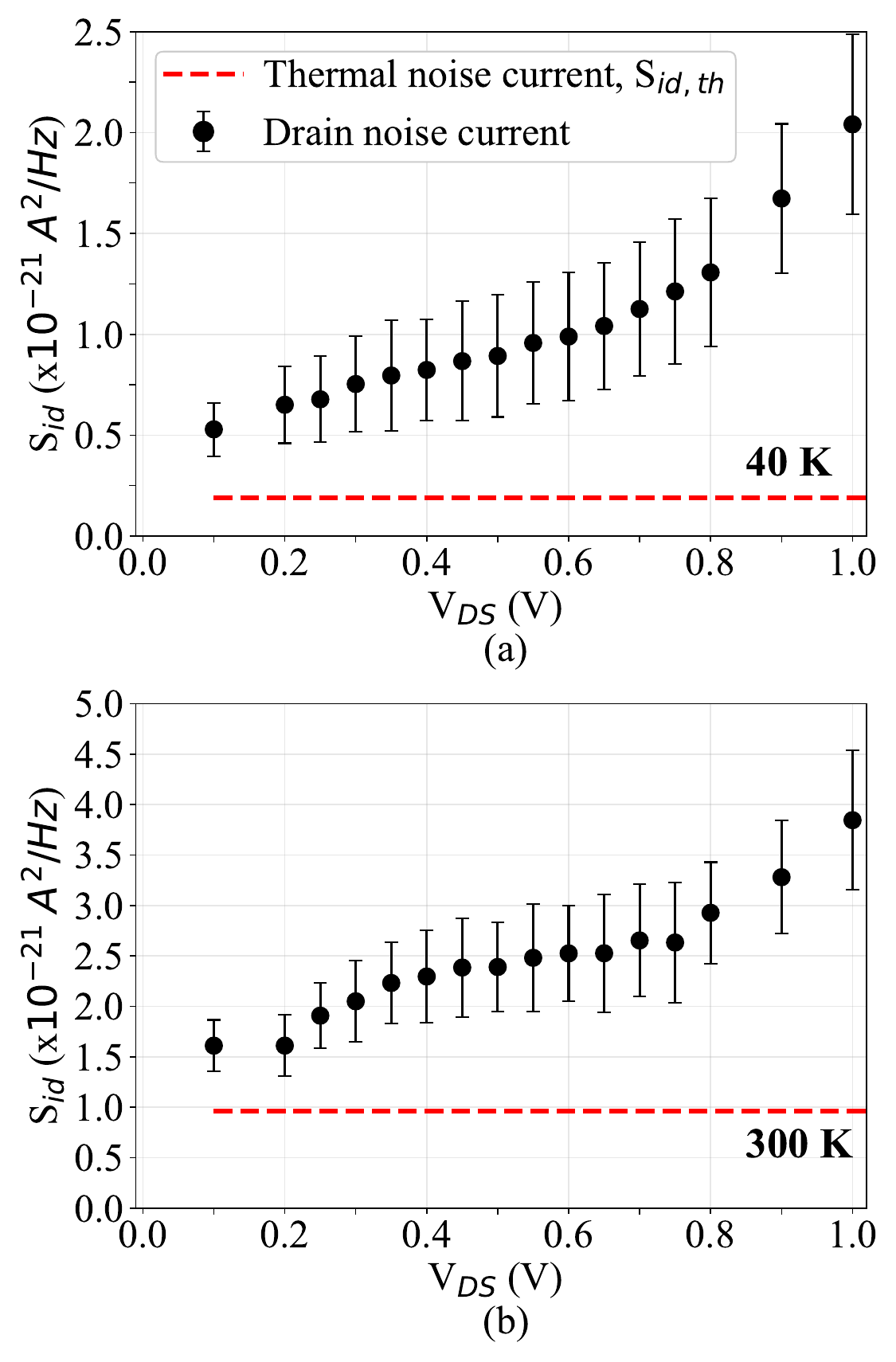}
{
\phantomsubcaption\label{fig:Td_Vds_GaAs_40K}
\phantomsubcaption\label{fig:Td_Vds_GaAs_300K}
}
\caption{ Extracted SSM $S_{id}$ (black circles) and computed channel thermal noise PSD $S_{th}$  (red dashed line) versus $V_{DS}$ at 40 K (a) and 300 K (b). The gate voltage, $V_{GS}$ was $-136$ mV at $T_{ph} = 40$ K, and $-226$ mV at $T_{ph} = 300$ K.}
\label{fig:Td_Te_RST_GaAs}
\end{figure}

We now discuss what physical mechanisms could account for the observed trends of drain noise PSD. Drain noise in HEMTs, like other field effect transistors, is expected to originate at least in part from thermal noise of the channel conductance. The relevant portion of the channel of interest for its noise contribution is that under the gate recess, as the other access resistances have been already taken into account in the SSM. The thermal noise of the channel subject to gate and drain biases is given by  $S_{id,th} = 4k_{B}g_{dso}T_{ph} \gamma$, where, $k_{B}$ is the Boltzmann constant, $g_{dso}$ is the channel conductance at $V_{DS} = 0$ V, and  $\gamma$ is a factor that accounts for the variation in carrier density along the channel due to the combined effect of the gate-source and drain-source potentials \cite{van_der_ziel_thermal_1983, channel_noise_MOSFETs_A, smit_rf-noise_2014}. In the linear region, $\gamma = 1$, while in saturation its values depend on the channel length; in the long-channel limit $\gamma = 2/3$ using the gradual-channel approximation, while in short-channel devices $\gamma$ varies between 1 and 2 due to short-channel effects such as velocity saturation and channel-length modulation \cite{Thermal_model_CMOS, channel_noise_MOSFETs_A, gamma_factor_less_2}. $S_{id,th}$ is also affected by high-field effects such as non-equilibrium carrier heating, meaning that electrons are heated above the lattice temperature. Although the higher electron temperature increases $S_{id}$, the high-field conductance is generally lower than the low-field value due to the decrease in mobility with increasing field. In addition, the carrier density in the pinch-off region is at least an order of magnitude lower than that of the unperturbed channel \cite{schwierz_modern_2002}. 


Considering these competing trends, various studies  have concluded that the noise from the pinch-off region is negligible \cite{channel_noise_MOSFETs_A, Vsat_and_Hot_el_effects, noise_Saturation_A}. This conclusion is compatible with experimental measurements of microwave noise in InGaAs quantum wells, which show that the current PSD decreases with increasing field \cite[Fig.~16.7]{hartnagel_microwave_2001}.
Therefore, an overestimate for the channel thermal noise may be obtained by taking $S_{id,th} = 4k_{B}g_{dso}T_{ph}$,  neglecting the decrease in noise associated with the formation of the pinch-off region at high $V_{DS}$. $g_{dso}$ is obtained from the slope of the I-V characteristics at $V_{DS}=0$ V. 

\begin{figure}
\centering
\includegraphics[width=0.95\columnwidth]{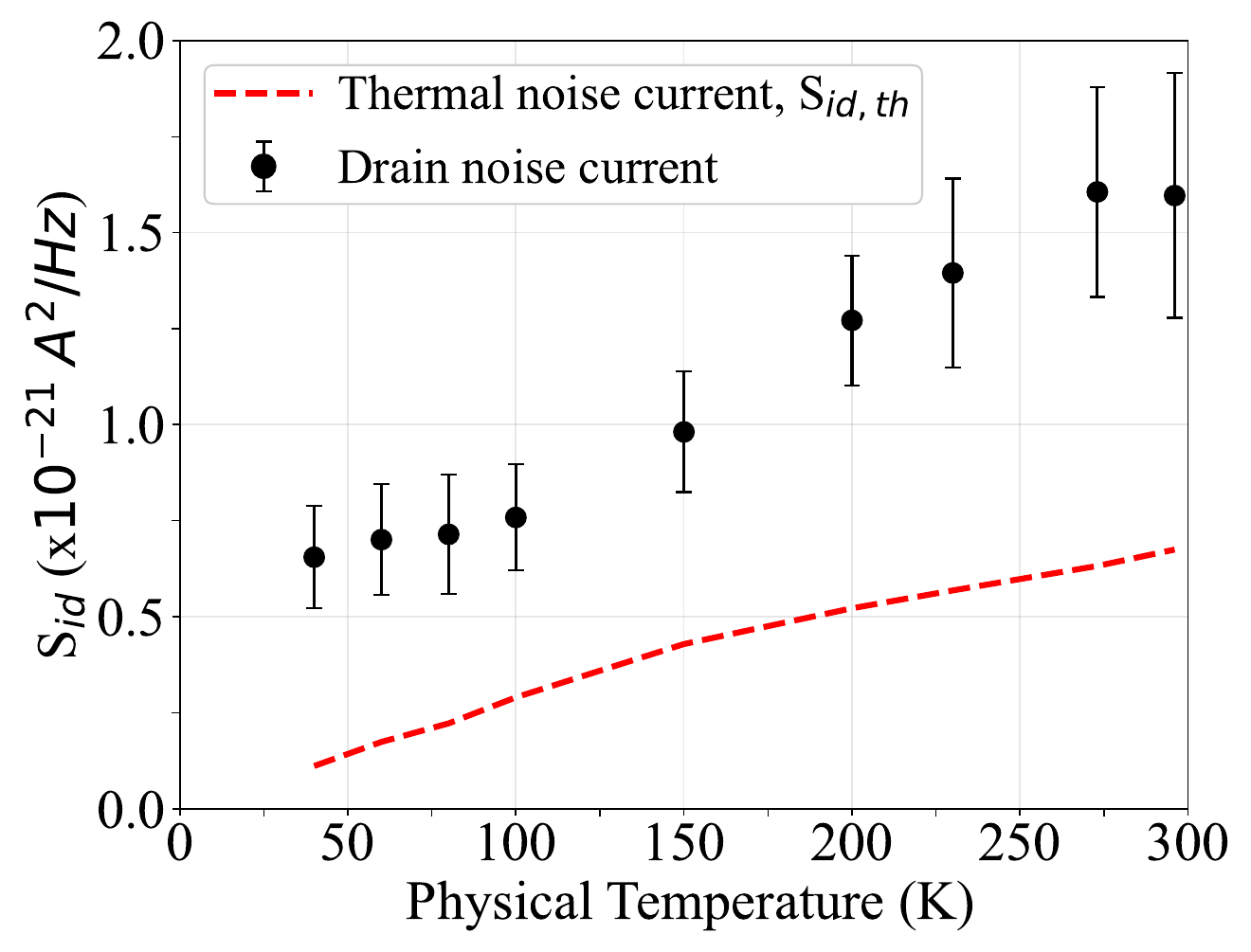}
\caption{ Extracted SSM drain noise PSD $S_{id}$ (black circles) versus physical temperature $T_{ph}$ and computed thermal noise PSD $S_{id,th}$ (red dashed line) at constant $V_{DS} = 0.6$ V and $I_{DS} = 10$ mA (computed from data reported in \cite{bekari}).}
\label{fig:Td_Te_RST_Vs_Tph_GaAs}
\end{figure}

In Figs.~\ref{fig:Td_Vds_GaAs_40K} and \ref{fig:Td_Vds_GaAs_300K}, the computed $S_{id,th}$ are shown at 40 K and 300 K, respectively. The measured $S_{id}$ exceeds the channel thermal noise magnitude, corresponding to $\gamma$ values of $\sim 3- 6$ at 40 K and $\sim 2-3$ at 300 K. However, theories of channel thermal noise including short-channel effects generally predict $\gamma \lesssim 1.5 $.  \cite{channel_noise_MOSFETs_A, Thermal_model_CMOS}
Fig.~\ref{fig:Td_Te_RST_Vs_Tph_GaAs} shows the $T_{ph}$ dependence of $S_{id}$ and $S_{id,th}$ at constant $V_{DS} = 0.6$ V and $I_{DS} = 10$ mA. For the calculation of $S_{id, th}$,  $g_{dso}$  was derived from the I-V curves at each $T_{ph}$ and was corrected for the access resistances following \cite{smit_rf-noise_2014, int_extr_conductance}. While $S_{id,th}$ qualitatively accounts for the temperature-dependence of $S_{id}$, the measured trend plateaus below 100 K while $S_{id,th}$ tends to zero with decreasing temperature. Additionally, the magnitude of $S_{id,th}$ is $\sim$50-70\% lower than the measured values. Considering the totality of the observations, the analysis suggests that another mechanism beyond channel thermal noise contributes to drain noise in HEMTs.


\section{ Discussion} \label{discussions}

We now discuss potential physical origins of the additional drain noise source. Suppressed shot noise can arise if electrons propagate quasi-ballistically in the channel. However, ultrafast optical studies have found that photoexcited electrons in quantum wells thermalize within 200 fs, implying an electron-electron collision time on the order of 10 fs.  \cite[Sec.~4C3b]{shah1986}. Considering a drift velocity of $\sim 5 \times 10^7$ cm s\textsuperscript{-1} \cite{rodilla_2013}, these values yield a mean free path of around 5-10 nm. This value is much shorter the gate length or source-drain separation of even highly-scaled HEMTs, implying that suppressed shot noise is not likely to be a relevant noise mechanism in HEMTs.

Another theory for drain noise based on real-space transfer of hot electrons from the channel to barrier films was developed following the conclusions of prior studies of microwave noise in quantum wells \cite{esho_theory_2022}. Specifically, in earlier experimental works the microwave noise tempjusterature was found to be affected by alterations to the quantum well potential even with the channel alloy composition held constant, suggesting that the heterojunction potential confining the channel electrons played a role in the noise mechanism \cite{aninkeviius_1993, aninkevicius_1994}. This observation is compatible with the real-space transfer mechanism in which hot electrons thermionically emit over the confining barrier potential to create partition noise. \cite{esho_theory_2022} derived an expression for $S_{id}$ in which noise arises due to RST; however, this work did not consider the contribution of channel thermal noise.

Here, we suggest that drain noise could be attributed to a combination of channel thermal noise and RST: $S_{id} = S_{th,id} + S_{RST}$.  The $S_{RST}$ is defined in \cite[Eq.~(3)]{esho_theory_2022} and depends exponentially on the peak electron temperature, the conduction band discontinuity ($\Delta E_{c}$) at the channel/barrier heterojunction and the overdrive voltage defined as $V_{ov} = V_{gs} - V_{th}$, where $V_{th}$ the threshold voltage. In \cite{esho_theory_2022}, the peak electron temperature was assumed to vary linearly with the physical temperature. However, Monte Carlo studies \cite{Fischetti} and the weak temperature-dependence of high-field velocity characteristics in many semiconductors indicate that the peak electron temperature in HEMTs should  exhibit a weak dependence on lattice temperature. In this case, RST noise would be predicted to exhibit only a weak dependence on physical temperature as well. This conclusion differs from that originally reported in \cite{esho_theory_2022}.

Qualitatively, the trend of $S_{id}$ versus $T_{ph}$ in Fig.~\ref{fig:Td_Te_RST_Vs_Tph_GaAs} can be explained if the thermal noise is responsible for the primary $T_{ph}$ dependence, with another mechanism with weaker temperature dependence accounting for the overall trend and magnitude. Using the numerical values given in \cite{esho_theory_2022} and assuming $V_{ov} = 43$ mV from the measurements in this work, we compute $S_{id, RST} \sim 3 \times 10^{-22}$ A$^2$/Hz over the range of temperatures studied, which is of the right order of magnitude to explain the discrepancy. However, we caution that this estimate is only valid to within a factor of 2-3 owing to uncertainties in the parameters described in \cite{esho_theory_2022}. 

The role of RST in drain noise could be further confirmed by examining how the microwave noise temperature trends with bias and temperature are affected by changes in quantum confinement of channel electrons, for instance by altering the conduction band offset while not varying the channel composition so as to avoid confounding effects from impact ionization, for example. This modification can be accomplished by increasing the Al composition of the barrier \cite{huang_measurement_1992}. Strained-barrier HEMTs with 
Al composition of over $55$\% have been reported \cite{strained_barrier_A}, and so such a study appears feasible.

\begin{figure}
\centering
\includegraphics[width=0.82\columnwidth]{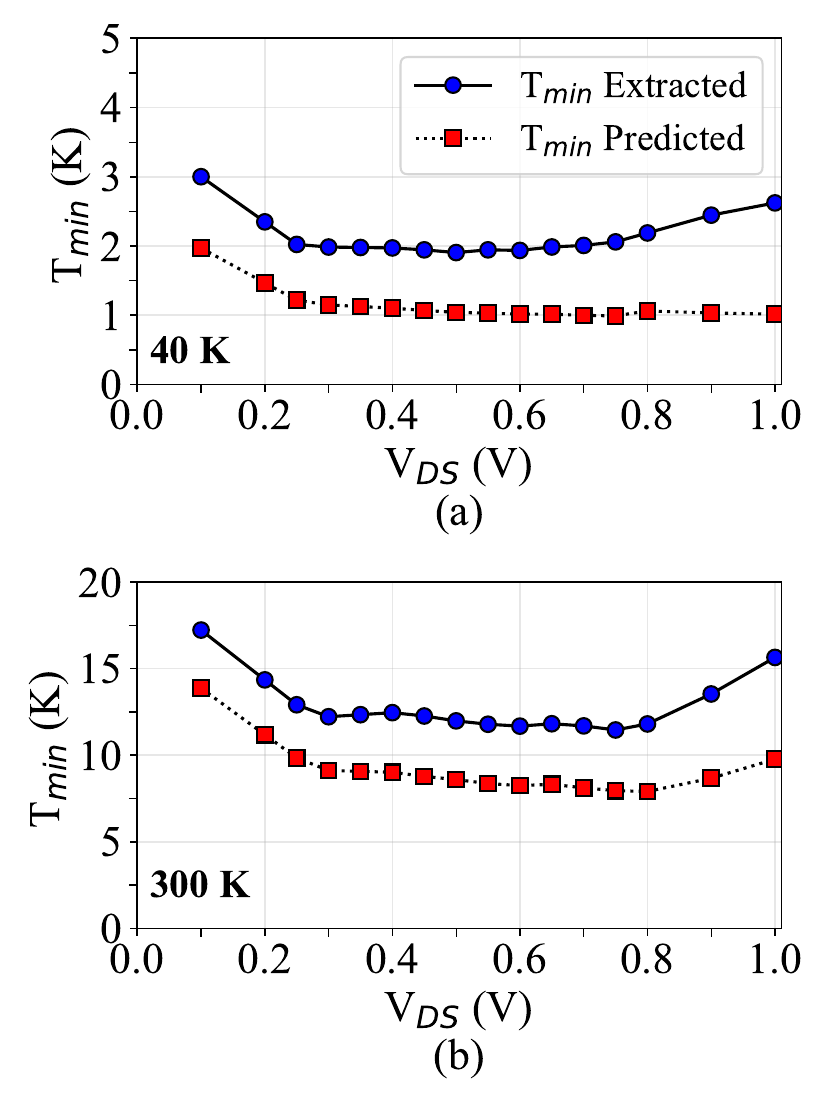}
{
\phantomsubcaption\label{fig:T50_Tpred_Tph}
\phantomsubcaption\label{fig:T50_Tpred_Vds_40K}
\phantomsubcaption\label{fig:T50_Tpred_Vds_300K}
}
\caption{ Extracted $T_{min}$ \cite{bekari} and predicted $T_{min}$ with only channel thermal noise versus $V_{DS}$ at 40 K (a) and 300 K (b) and constant $V_{GS}=-136$ mV and $V_{GS}=-226$ mV, respectively.   
The predicted $T_{min}$ is obtained by replacing by setting $S_{id}= S_{th}$ in the noise model. All the above data are at the frequency of 6 GHz. The $V_{GS}$ in (a) and (b) were selected so that at both physical temperatures, $V_{DS}=0.8$ V and $I_{ds}=20$ mA.}
\label{fig:T50_Meas_Vs_Predicted}
\end{figure}

Finally, we use our noise model to estimate the magnitude of improvement in $T_{min}$ if the channel noise were only due to thermal noise. In Ref.~\ref{fig:T50_Meas_Vs_Predicted}, we plot the extracted $T_{min}$, obtained from the noise model using S-parameter and $T_{50}$ data versus $V_{DS}$ at $T_{ph}$ of 40 K and 300 K, and frequency of 6 GHz. In this plot, we also show the predicted trend of $T_{min}$ if drain noise was due solely to channel thermal noise. We observe that the minimum $T_{min}$ could be improved by $\sim 50$\% and $\sim 30$\%  at 40 K and 300 K, respectively. This result implies that if the hot-electron noise observed in $S_{id}$ originates from RST, then significant improvements in cryogenic and room temperature noise performance may be possible by engineering the quantum well for improved quantum confinement.

\section{Conclusion} \label{conclusion}
We have characterized the $S$-parameters and the microwave noise temperature of InGaAs mHEMTs at  40 K and 300 K and various $V_{DS}$. The extracted drain noise is found to exceed that expected from channel thermal noise by a factor of $\sim$2-6, suggesting that an additional mechanism contributes to drain noise. Based on prior studies of microwave noise in quantum wells, we hypothesize that this noise mechanism is real-space transfer noise. We suggest approaches to further test this hypothesis. Finally, we compute that improvements in minimum cryogenic noise temperature of up to 50\% can be achieved if the hot-electron noise is suppressed.

\section*{Acknowledgments}
The authors thank Jan Grahn, Pekka Kangaslahti, Jacob Kooi,  Junjie Li, Jun Shi and Sander Weinreb for useful discussions. 

\ifCLASSOPTIONcaptionsoff
  \newpage
\fi

\bibliographystyle{IEEEtran}
\bibliography{IEEEabrv,main}
%
\newpage

\end{document}